\documentstyle[twoside,psfig]{article}

\catcode`\@=11
\long\def\@makefntext#1{
\protect\noindent \hbox to 3.2pt {\hskip-.9pt  
$^{{\eightrm\@thefnmark}}$\hfil}#1\hfill}		

\def\@makefnmark{\hbox to 0pt{$^{\@thefnmark}$\hss}}	
	
\def\ps@myheadings{\let\@mkboth\@gobbletwo
\def\@oddhead{\hbox{}
\rightmark\hfil\eightrm\thepage}   
\def\@oddfoot{}\def\@evenhead{\eightrm\thepage\hfil
\leftmark\hbox{}}\def\@evenfoot{}
\def\sectionmark##1{}\def\subsectionmark##1{}}

\oddsidemargin=\evensidemargin
\newcounter{sectionc}\newcounter{subsectionc}\newcounter{subsubsectionc}
\renewcommand{\section}[1] {\vspace{12pt}\addtocounter{sectionc}{1} 
\setcounter{subsectionc}{0}\setcounter{subsubsectionc}{0}\noindent 
	{\tenbf\thesectionc. #1}\par\vspace{5pt}}
\renewcommand{\subsection}[1] {\vspace{12pt}\addtocounter{subsectionc}{1} 
	\setcounter{subsubsectionc}{0}\noindent 
	{\bf\thesectionc.\thesubsectionc. {\kern1pt \bfit #1}}\par\vspace{5pt}}
\renewcommand{\subsubsection}[1] {\vspace{12pt}\addtocounter{subsubsectionc}{1}
	\noindent{\tenrm\thesectionc.\thesubsectionc.\thesubsubsectionc.
	{\kern1pt \tenit #1}}\par\vspace{5pt}}
\newcommand{\nonumsection}[1] {\vspace{12pt}\noindent{\tenbf #1}
	\par\vspace{5pt}}

\newcounter{appendixc}
\newcounter{subappendixc}[appendixc]
\newcounter{subsubappendixc}[subappendixc]
\renewcommand{\thesubappendixc}{\Alph{appendixc}.\arabic{subappendixc}}
\renewcommand{\thesubsubappendixc}
	{\Alph{appendixc}.\arabic{subappendixc}.\arabic{subsubappendixc}}

\renewcommand{\appendix}[1] {\vspace{12pt}
        \refstepcounter{appendixc}
        \setcounter{figure}{0}
        \setcounter{table}{0}
        \setcounter{lemma}{0}
        \setcounter{theorem}{0}
        \setcounter{corollary}{0}
        \setcounter{definition}{0}
        \setcounter{equation}{0}
        \renewcommand{\thefigure}{\Alph{appendixc}.\arabic{figure}}
        \renewcommand{\thetable}{\Alph{appendixc}.\arabic{table}}
        \renewcommand{\theappendixc}{\Alph{appendixc}}
        \renewcommand{\thelemma}{\Alph{appendixc}.\arabic{lemma}}
        \renewcommand{\thetheorem}{\Alph{appendixc}.\arabic{theorem}}
        \renewcommand{\thedefinition}{\Alph{appendixc}.\arabic{definition}}
        \renewcommand{\thecorollary}{\Alph{appendixc}.\arabic{corollary}}
        \renewcommand{\theequation}{\Alph{appendixc}.\arabic{equation}}
        \noindent{\tenbf Appendix \theappendixc #1}\par\vspace{5pt}}
\newcommand{\subappendix}[1] {\vspace{12pt}
        \refstepcounter{subappendixc}
        \noindent{\bf Appendix \thesubappendixc. {\kern1pt \bfit #1}}
	\par\vspace{5pt}}
\newcommand{\subsubappendix}[1] {\vspace{12pt}
        \refstepcounter{subsubappendixc}
        \noindent{\rm Appendix \thesubsubappendixc. {\kern1pt \tenit #1}}
	\par\vspace{5pt}}

\newcommand{\textlineskip}{\baselineskip=13pt}
\newcommand{\smalllineskip}{\baselineskip=10pt}

\def\eightcirc{
\begin{picture}(0,0)
\put(4.4,1.8){\circle{6.5}}
\end{picture}}
\def\eightcopyright{\eightcirc\kern2.7pt\hbox{\eightrm c}} 

\newcommand{\copyrightheading}[1]
	{\vspace*{-2.5cm}\smalllineskip{\flushleft
	{\footnotesize Modern Physics Letters A, #1}\\
	{\footnotesize $\eightcopyright$\, World Scientific Publishing
	 Company}\\
	 }}

\newcommand{\publisher}[2]{{\begin{center}\footnotesize\smalllineskip 
	Received #1\\
	Revised #2
	\end{center}
	}}

\def\abstracts#1#2#3{{
	\centering{\begin{minipage}{4.5in}\footnotesize\baselineskip=10pt
	\parindent=0pt #1\par 
	\parindent=15pt #2\par
	\parindent=15pt #3
	\end{minipage}}\par}} 

\newcommand{\bibit}{\nineit}
\newcommand{\bibbf}{\ninebf}
\renewenvironment{thebibliography}[1]
	{\frenchspacing
	 \ninerm\baselineskip=11pt
	 \begin{list}{\arabic{enumi}.}
        {\usecounter{enumi}\setlength{\parsep}{0pt}     
	 \setlength{\leftmargin 12.7pt}{\rightmargin 0pt} 
         \setlength{\itemsep}{0pt} \settowidth
	{\labelwidth}{#1.}\sloppy}}{\end{list}}

\newcounter{itemlistc}
\newcounter{romanlistc}
\newcounter{alphlistc}
\newcounter{arabiclistc}

\newcommand{\fcaption}[1]{
        \refstepcounter{figure}
        \setbox\@tempboxa = \hbox{\footnotesize Fig.~\thefigure. #1}
        \ifdim \wd\@tempboxa > 5in
           {\begin{center}
        \parbox{5in}{\footnotesize\smalllineskip Fig.~\thefigure. #1}
            \end{center}}
        \else
             {\begin{center}
             {\footnotesize Fig.~\thefigure. #1}
              \end{center}}
        \fi}

\newcommand{\tcaption}[1]{
        \refstepcounter{table}
        \setbox\@tempboxa = \hbox{\footnotesize Table~\thetable. #1}
        \ifdim \wd\@tempboxa > 5in
           {\begin{center}
        \parbox{5in}{\footnotesize\smalllineskip Table~\thetable. #1}
            \end{center}}
        \else
             {\begin{center}
             {\footnotesize Table~\thetable. #1}
              \end{center}}
        \fi}

\def\@citex[#1]#2{\if@filesw\immediate\write\@auxout
	{\string\citation{#2}}\fi
\def\@citea{}\@cite{\@for\@citeb:=#2\do
	{\@citea\def\@citea{,}\@ifundefined
	{b@\@citeb}{{\bf ?}\@warning
	{Citation `\@citeb' on page \thepage \space undefined}}
	{\csname b@\@citeb\endcsname}}}{#1}}

\newif\if@cghi
\def\cite{\@cghitrue\@ifnextchar [{\@tempswatrue
	\@citex}{\@tempswafalse\@citex[]}}
\def\citelow{\@cghifalse\@ifnextchar [{\@tempswatrue
	\@citex}{\@tempswafalse\@citex[]}}
\def\@cite#1#2{{$\null^{#1}$\if@tempswa\typeout
	{IJCGA warning: optional citation argument 
	ignored: `#2'} \fi}}

\def\pmb#1{\setbox0=\hbox{#1}
	\kern-.025em\copy0\kern-\wd0
	\kern.05em\copy0\kern-\wd0
	\kern-.025em\raise.0433em\box0}

\def\fnt#1#2{\footnotetext{\kern-.3em
	{$^{\mbox{\scriptsize #1}}$}{#2}}}

\def\fpage#1{\begingroup
\voffset=.3in
\thispagestyle{empty}\begin{table}[b]\centerline{\footnotesize #1}
	\end{table}\endgroup}

\def\runninghead#1#2{\pagestyle{myheadings}
\markboth{{\protect\footnotesize\it{\quad #1}}\hfill}
{\hfill{\protect\footnotesize\it{#2\quad}}}}
\font\tenrm=cmr10
\font\tenit=cmti10 
\font\tenbf=cmbx10
\font\bfit=cmbxti10 at 10pt
\font\ninerm=cmr9
\font\nineit=cmti9
\font\ninebf=cmbx9
\font\eightrm=cmr8

\def\qed{\hbox{${\vcenter{\vbox{			
   \hrule height 0.4pt\hbox{\vrule width 0.4pt height 6pt
   \kern5pt\vrule width 0.4pt}\hrule height 0.4pt}}}$}}

\begin{document}
\setlength{\textheight}{7.7truein}  

\runninghead{D. V. Ahluwalia, Y. Liu, I. Stancu}
{CP violation in neutrino oscillations   $\ldots$}

\normalsize\textlineskip
\thispagestyle{empty}
\setcounter{page}{1}

\copyrightheading{Vol. 17, No. 1 (2002) pp.13-21.}		

\vspace*{0.88truein}

\def\beq{\begin{eqnarray}}
\def\eeq{\end{eqnarray}}



\def\@citex[#1]#2{\if@filesw\immediate\write\@auxout
	{\string\citation{#2}}\fi	
\def\@citea{}\@cite{\@for\@citeb:=#2\do		
	{\@citea\def\@citea{,}\@ifundefined	
	{b@\@citeb}{{\bf ?}\@warning 
	{Citation `\@citeb' on page \thepage \space undefined}} 
	{\csname b@\@citeb\endcsname}}}{#1}} 
 \newif\if@cghi 
\def\cite{\@cghitrue\@ifnextchar [{\@tempswatrue 
	\@citex}{\@tempswafalse\@citex[]}} 
\def\citelow{\@cghifalse\@ifnextchar [{\@tempswatrue 
	\@citex}{\@tempswafalse\@citex[]}} 
\def\@cite#1#2{{[{#1}]\if@tempswa\typeout           
	{WSPC warning: optional citation argument  
	ignored: `#2'} \fi}} 
%
\def\@refcitex[#1]#2{\if@filesw\immediate\write\@auxout  
	{\string\citation{#2}}\fi 
\def\@citea{}\@refcite{\@for\@citeb:=#2\do 
	{\@citea\def\@citea{, }\@ifundefined 
	{b@\@citeb}{{\bf ?}\@warning 
	{Citation `\@citeb' on page \thepage \space undefined}} 
	\hbox{\csname b@\@citeb\endcsname}}}{#1}} 
 \def\@refcite#1#2{{[{#1}]\if@tempswa\typeout     
        {WSPC warning: optional citation argument 
	ignored: `#2'} \fi}} 
 \def\refcite{\@ifnextchar[{\@tempswatrue 
	\@refcitex}{\@tempswafalse\@refcitex[]}} 
\def\dfrac#1#2{{\displaystyle{#1\over#2}}}
\def\tfrac#1#2{{\textstyle{#1\over#2}}}


\fpage{1}
\centerline{\bf CP VIOLATION IN NEUTRINO OSCILLATIONS
AND L/E FLATNESS }
\baselineskip=13pt
\centerline{\bf  OF THE E-LIKE EVENT RATIO AT SUPER-KAMIOKANDE} 
\vspace*{0.37truein}
\centerline{\footnotesize D. V. AHLUWALIA\footnote{E-mail: 
ahluwalia@phases.reduaz.mx, http://phases.reduaz.mx}}
\baselineskip=12pt
\centerline{\footnotesize\it Facul. de Fisica de la UAZ, Ap. Postal C-600}
\baselineskip=10pt
\centerline{\footnotesize\it Zacatecas, ZAC 98062, Mexico}
\vspace*{10pt}

\centerline{\footnotesize Y. LIU\footnote{E-mail:  yong@isn.in2p3.fr}}
\baselineskip=12pt
\centerline{\footnotesize\it Istitut des Sciences Nucleaires, IN2P3,
}
\baselineskip=10pt
\centerline{\footnotesize\it 53 av. des Martyrs, 38026 Grenoble cedex, France}
\vspace*{0.225truein}

\centerline{\footnotesize I. STANCU\footnote{E-mail: istancu@bama.ua.edu}}
\baselineskip=12pt
\centerline{\footnotesize\it Department of Physics and Astronomy}
\baselineskip=10pt
\centerline{\footnotesize\it University of Alabama, Tuscaloosa, 
Alabama, AL 35487, USA}
\vspace*{0.225truein}

\publisher{(received date)}{(revised date)}

\vspace*{0.21truein}
\abstracts{
We show that if the presently observed $L/E$-flatness of the 
electron-like  event ratio in the Super-Kamiokande atmospheric 
neutrino data is confirmed then the indicated ratio must be {\em unity}. 
Further, it is found that once CP is violated the exact L/E flatness implies:
(a)  The CP-violating phase, in the standard parameterization,
is narrowed down to two possibilities $\pm \pi/2$,
and (b) The mixing between the second and the third
generations must be maximal.
With these results at hand, we argue that
a dedicated study of the  $L/E$-flatness of the electron-like event
ratio by Super-Kamiokande can serve as
an initial investigatory probe of CP violation in the neutrino sector.
The assumptions under which these 
results hold are explicitly stated.}{}{}

\setcounter{footnote}{0}

\section{Introduction}

\noindent
  The Super-Kamiokande data on the atmospheric neutrinos  have opened
  a new realm of physics research  \cite{kek}. The simplest interpretation
  of these data is flavor oscillations arising from
  neutrino being linear superposition of some underlying mass eigenstates.
  This circumstance not only takes us into the physics beyond the standard
  model of the high energy physics, but it also allows to probe various
  aspects of quantum gravity \cite{qg0,qg1,qg2,qg3,qg4}. 
As such much theoretical
  and experimental effort is being devoted to deciphering the nature
  of neutrino. Here, using a very specific aspect of the Super Kamiokande
  data, we shall analytically constrain the CP-violating neutrino
  oscillation mixing matrix. This would help the design of future
experiments,
  allow for more analytically-oriented theoretical research, and provide
  a new direction of research at the existing experimental facilities.

  This work joins the on-going research with the observation that
  as soon as the first results from the  Super-Kamiokande on atmospheric
  neutrinos became available, one of us emphasized that the L/E flatness
  noted in the {\em abstract\/} places a set of constraints on
  the neutrino oscillation mixing matrix \cite{ahluw}. However, in that,
  and our subsequent work \cite{isdva}, CP violation
  has been neglected. Apart from reasons of simplicity,
  there is no {\em a priori\/} reason to assume the absence of CP
  violation in the neutrino sector.
  In addition, the observed cosmological baryonic asymmetry may be
  deeply connected with a CP violation in the leptonic sector \cite{ls}.
  This becomes
  particularly important,
  as we shall comment below, if the neutrino-sector CP violation is
  affected by gravity.
  As such, here,
  we present a non-trivial generalization of the  constraints presented
  in the early work \cite{ahluw,isdva}
  to obtain
  a CP-violating bimaximal matrix for neutrino oscillations.\footnote{
  To avoid confusion, we note in advance that in this paper
  we distinguish
  between
  {\it flux\/} and {\it events\/}. The former refers to the number
  of particles of a given species that pass a unit area in a
  unit time, while the latter depends on the detector sensitivity and
  the relevant cross sections.
}

\section{Analytical constraints on the neutrino-oscillation mixing matrix}
\noindent
  To generalize the discussion of Refs. \refcite{ahluw,isdva}, 
we start from the
  probability formula of neutrino oscillations. As in the
  quark sector, when neutrinos
  have non-zero masses, their weak eigenstates may not
  coincide with the mass eigenstates, but may be linear superposition of the
mass eigenstates.
  The latter choice is precisely what is suggested by the existing
  data \refcite{kek,solar,LSND,KARMEN,K2K}. As such,
  in a phenomenology of neutrino oscillations,
  a flavor eigenstate of a neutrino is
  postulated to be a linear superposition of some underlying mass
  eigenstates
  \begin{equation}
  \mid \nu_{\alpha} \rangle = \sum_j U_{\alpha j} \mid
  \nu_j \rangle,
  \end{equation}
  where $U_{\alpha j}$ is an element of the mixing matrix
  with
  $ \alpha = e, \; \mu, \; {\rm or}\; \tau$ and $j=1, \;
  2, \; 3$
  in the framework of three generations. In the
  literature, $U$ is usually
  taken as the standard parameterization matrix
  \refcite{part}
  \begin{equation}
  V=
  \left(
  \begin{array}{ccc}
  c_{12} c_{13} & s_{12} c_{13} & s_{13}e^{-i
  \delta_{13} } \\
  -s_{12} c_{23}-c_{12} s_{23} s_{13} e^{i \delta_{13}}
  &
  c_{12} c_{23}-s_{12} s_{23} s_{13} e^{i \delta_{13}} &
  s_{23}c_{13}\\
  s_{12} s_{23}-c_{12} c_{23} s_{13} e^{i \delta_{13}} &
  -c_{12} s_{23}-s_{12} c_{23} s_{13} e^{i \delta_{13}}
  &
  c_{23}c_{13}
  \end{array}
  \right)
  \end{equation}
  multiplied by a phase matrix
  \begin{equation}
  P=
  \left(
  \begin{array}{ccc}
  1 & 0&0 \\
  0&e^{i \phi_2} & 0\\
  0 &0 &  e^{i \phi_3+\delta_{13}}
  \end{array}
  \right)
  \end{equation}
  if neutrinos are of the Majorana type.
  Here, $c_{ij}=\cos\theta_{ij}, \;
  s_{ij}=\sin\theta_{ij}$, and $\phi_2$  and $\phi_3$
  are the additional phases for Majorana neutrinos.
  Due to the un-observable
  effect of $P$ in flavor oscillation experiments, we will drop
  it in the discussion
  and simply equate the mixing matrix $U$ to $V$ in
  calculations that follow.
  Furthermore, $\theta_{12},\; \theta_{23},$ and $\theta_{13}$ in $U$
  can all be made  to lie in the
  first quadrant by an appropriate re-definition of the
  relevant fields.

  Assuming the underlying mass eigenstates to be relativistic in the
  observer's frame \refcite{ag},  the flavor-oscillation probability
  is given by \refcite{isdva,dick}
  \begin{eqnarray}
  \label{pp1}
  P(\nu_{\alpha} \stackrel{L}{ \rightarrow} \nu_{\beta})
  =\delta_{\alpha \beta} & - & 4 \sum_{j<k} {\rm Re}(
  U_{\alpha j} U_{\beta j}^{\ast} U_{\alpha k}^{\ast}
  U_{\beta k})\, {\sin}^2\left(\varphi_{jk}\right)
  \nonumber\\
  & + & 2 \sum_{j<k}  {\rm Im}(
  U_{\alpha j} U_{\beta j}^{\ast} U_{\alpha k}^{\ast}
  U_{\beta k})\,
  {\sin}\left(2\,\varphi_{jk}\right).
  \end{eqnarray}
  where $L$, measured in  meters, refers to the source-detector distance,
and
  the flavor-oscillation inducing
  kinematic phases $\varphi_{ij}$,
  are defined as
  \beq
  \varphi_{ij} = 1.27\,\Delta m^2_{ij}\frac{L}{E},
  \eeq
  where E (MeV) refers to the ``energy'' of the flavor state, and,
   $\Delta m^2_{ij}=m_i^2-m_j^2$, is the mass-squared
  difference of the underlying
  mass eigenstates and is measured in
  eV$^2$.

  For the CP conjugate channel, the CP-odd term, that
  is, the last term in Eq. (\ref{pp1}),
  changes sign. So,
  \begin{eqnarray}
  \label{pp2}
  P({\bar \nu_{\alpha}} \stackrel{L}{ \rightarrow} {\bar
  \nu_{\beta}})
  =\delta_{\alpha \beta} & - & 4 \sum_{j<k}  {\rm Re}(
  U_{\alpha j} U_{\beta j}^{\ast} U_{\alpha k}^{\ast}
  U_{\beta k})
  {\sin}^2\left(\varphi_{jk}\right) \nonumber\\
  & - & 2 \sum_{j<k}  {\rm Im}(
  U_{\alpha j} U_{\beta j}^{\ast} U_{\alpha k}^{\ast}
  U_{\beta k})
  {\sin}\left(2\,\varphi_{jk}\right).
  \end{eqnarray}
  Note that,
  all $ {\rm Im}(U_{\alpha j} U_.{\beta j}^{\ast}
  U_{\alpha k}^{\ast} U_{\beta k}) $
  with $\alpha \neq \beta $ and $ j \neq k $ take
  the same value
  $J_{CP}=c_{12}s_{12}c_{23}s_{23}c_{13}^2
  s_{13}s_{\delta}
  \;(s_{\delta}={\sin}\delta_{13},\; c_{\delta}={\cos}\delta_{13} )$,
  which is the measure of CP violation \refcite{ajar}.

  The Super-Kamiokande measured ratio, ${\cal R}_e$, of the
  electron-like events is defined as
  \begin{equation}
  \label{zz3}
  {\cal R}_e=\frac{N_e^\prime + N_{\bar e}^\prime}{N_e +
  N_{\bar e}}.
  \end{equation}
  where $N_e$ and $N_{\bar e}$ are the numbers of {\em predicted\/} $\nu_e$
  and $\bar\nu_e$ events in the absence of neutrino oscillations,
  whereas the primed quantities are the corresponding numbers of
  {\em observed} events, allowing for the presence of neutrino
  oscillations.

  If at the top of atmosphere, i.e. the ``source,'' the number of
  $\nu_e$ $(\bar\nu_e)$ and $\nu_\mu$ $(\bar\nu_\mu)$ are
  $N_{\nu_e} (N_{\bar \nu_e}) $
  and $N_{\nu_{\mu}} (N_{\bar \nu_{\mu}})$ respectively,
  while the
  cross-sections for $\nu_e$ and ${\bar \nu_e}$ are
  $\sigma_{\nu_e}$ and
  $\sigma_{\bar \nu_e}$; then we obtain the following set of
  event predictions for the detector:

\beq
  \label{zz1}
  N_e&=&N_{\nu_e}\sigma_{\nu_e}\\
  N_{\bar e}&=&N_{\bar \nu_e} \sigma_{\bar \nu_e}\\
  N_e^\prime&=&N_{\nu_e} P(\nu_e \stackrel{L}{
  \rightarrow} \nu_e) \sigma_{\nu_e}
  +N_{\nu_\mu} P(\nu_\mu \stackrel{L}{ \rightarrow}
  \nu_e) \sigma_{\nu_e}\\
  \label{zz2}
   N_{\bar e}^\prime&=&N_{\bar \nu_e} P({\bar \nu_{ e}}
  \stackrel{L}{ \rightarrow}{\bar  \nu_{ e}})
  \sigma_{\bar \nu_e}
  +N_{\bar \nu_\mu} P({\bar \nu_\mu}
   \stackrel{L}{ \rightarrow} {\bar \nu_e}) \sigma_{\bar
  \nu_e}.
  \eeq
  The first two equations correspond to absence of flavor oscillations,
  and the last two equations incorporate effects of flavor oscillations
  of neutrinos.

  Now, inserting Eqs. (\ref{zz1}-\ref{zz2}) into Eq. (\ref{zz3}),
  and taking note of the fact that
  due to $CPT$ symmetry,
  $$
  P(\nu_e \stackrel{L}{ \rightarrow} \nu_e)
  =P({\bar \nu_{ e}}\stackrel{L}{ \rightarrow} {\bar
  \nu_{ e}}),
  $$
  we arrive at
  \begin{equation}
  {\cal R}_e - P(\nu_e \stackrel{L}{ \rightarrow}
  \nu_e)=
  \frac{N_{\nu_\mu} P(\nu_\mu \stackrel{L}{ \rightarrow}
  \nu_e) \sigma_{\nu_e}
  +N_{\bar \nu_\mu} P({\bar \nu_\mu} \stackrel{L}{
  \rightarrow} {\bar \nu_e})
   \sigma_{\bar \nu_e}}{N_{\nu_e} \sigma_{\nu_e}
  +N_{\bar \nu_e}  \sigma_{\bar \nu_e}}.
  \end{equation}
  Finally, on defining
  \begin{eqnarray}
  \label{differe}
  \frac{N_{\bar \nu_e}}{N_{\nu_e}}=x, \hspace{0.5cm}
  \frac{N_{\bar \nu_\mu}}{N_{\nu_\mu}}=y \nonumber\\
  \frac{\sigma_{\bar \nu_e}}{\sigma_{\nu_e}}=\lambda,
  \hspace{0.5cm}
  \frac{N_{ \nu_\mu}}{N_{\nu_e}}=r,
  \end{eqnarray}
  it is easy to show that
  \begin{equation}
  {\cal R}_e - P(\nu_e \stackrel{L}{ \rightarrow} \nu_e)
  = \frac{r}{1+\lambda x} ( P(\nu_{\mu} \stackrel{L}{
  \rightarrow} \nu_{e})
  + \lambda y P({\bar \nu_{ \mu}} \stackrel{L}{
  \rightarrow} {\bar \nu_{ e}})).
  \end{equation}

  Now, substituting Eqs. (\ref{pp1},\ref{pp2}) into the above
  equation, and after simplifying,
  we obtain
\beq
  \label{r0}
  && \left\{ \left\vert U_{e1} \right\vert^2  \left\vert U_{e2}\right\vert^2
 +
  r \frac{1+\lambda y}{1+\lambda x} {\rm Re}( U_{\mu 1}
  U_{e 1}^{\ast}
  U_{\mu 2}^{\ast} U_{e2} )  \right\}\,
  {\sin}^2 \left(\varphi_{12}\right) \nonumber \\
  &+& \left\{  \left\vert U_{e1}\right\vert^2
  \left\vert U_{e3} \right\vert^2 +
  r  \frac{1+\lambda y}{1+\lambda x}{\rm Re}( U_{\mu 1}
  U_{e 1}^{\ast}
  U_{\mu 3}^{\ast} U_{e3} ) \right\}\,
  {\sin}^2 \left(\varphi_{13}\right)  \nonumber\\
  &+& \left\{  \left\vert U_{e 2}\right\vert^2
  \left\vert U_{e 3} \right\vert^2 +
  r  \frac{1+\lambda y}{1+\lambda x}{\rm Re}( U_{\mu 2}
  U_{e 2}^{\ast}
  U_{\mu 3}^{\ast} U_{e 3} )  \right\}\,
  {\sin}^2 \left(\varphi_{23}\right)  \nonumber\\
  \quad\quad\quad & -& \frac{r}{2}  \frac{1-\lambda y}{1+\lambda x}
  J_{CP} \left[ \sin \left(2\,\varphi_{12}\right) +
   \sin\left(2\,\varphi_{13}\right) + {\sin}\left(2\,\varphi_{23}\right)
  \right]
   \nonumber\\
  &=& \frac{1}{4} \left(1 - {\cal R}_e \right).
  \eeq
  It is worth  noting that  in case $x=y$ and $J_{CP}=0$,
  i.e., if the ratio of the numbers of
  $\bar\nu_e$  to $\nu_e$ equals
  the ratio of the numbers of
  $\bar\nu_\mu$  to $\nu_\mu$ at the source
  , and if there is
  no CP violation in the neutrino sector, Eq. ({\ref{r0}}) looses
  dependence on
  the neutrino and anti-neutrino cross sections.

  In order that Eq. (\ref{r0}) holds for all values of
   $L/E$  we must impose the constraints:\footnote{The Super-Kamiokande 
data spans roughly five orders of magnitude in $L/E$. However, as a 
mathematical theorem, it can be shown that if ${\cal R}_e$ carries
an $L/E$ independence over a finite range of $L/E$ then it must be 
so over the entire range of $L/E$.}

  \begin{equation}
  \label{xj}
  \frac{r}{2} \frac{1-\lambda y}{1+\lambda x} J_{CP}=0
  \end{equation}
  and
  \begin{eqnarray}
  \label{f1}
  \left\vert U_{e1} \right\vert^2 \left\vert U_{e2}\right\vert^2
  + r  \frac{1+\lambda y}{1+\lambda x}
   {\rm Re}( U_{\mu 1} U_{e 1}^{\ast} U_{\mu 2}^{\ast}
  U_{e2} )&=&0\\
  \left\vert U_{e1}\right\vert^2 \left\vert U_{e3}\right\vert^2 +
  r  \frac{1+\lambda y}{1+\lambda x}
   {\rm Re}( U_{\mu 1} U_{e 1}^{\ast} U_{\mu 3}^{\ast}
  U_{e3} ) &=&0\\
  \label{f3}
  \left\vert U_{e 2}\right\vert^2 \left\vert U_{e 3} \right\vert^2 +
  r  \frac{1+\lambda y}{1+\lambda x}
   {\rm Re}( U_{\mu 2} U_{e 2}^{\ast} U_{\mu 3}^{\ast}
  U_{e 3} ) &=&0.
  \end{eqnarray}

\section{The constrained CP-violating matrix} 

\noindent
  Since Eq. (\ref{r0}) holds for any value of
  $L/E$, we are also free to set
  $L/E=0$. This yields:
  \begin{equation}
  {\cal R}_e = 1.
  \end{equation}
  Although we invoke the Super-Kamiokande observed flatness for
 ${\cal R}_e$ from the beginning, we did  {\em not\/} refer to a specific
 value of  ${\cal R}_e$.
  The present analysis {\em predicts\/} ${\cal R}_e$ to be unity.
  This circumstance is in sharp contrast to the framework of
  references \refcite{ahluw,isdva} where one assumes both the
  indicated flatness
  and the value unity for ${\cal R}_e$.

  Furthermore,  Eq. (\ref{xj}) requires that  $J_{CP}=0$ and/or
  $\lambda y=1$. We consider each of these cases in turn.

\subsection{{$J_{CP}=0$} Case:}

The constraints (\ref{xj}-\ref{f3}), after some algebraic manipulations,
reduce to:

  \begin{eqnarray}
  \label{111}
  c_{12} s_{12} c_{13}^2 +  r  \frac{1+\lambda y}{1+\lambda x} \{
  c_{12} s_{12} (s_{23}^2 s_{13}^2 - c_{23}^2)
  +(s_{12}^2-c_{12}^2) c_{23} s_{23} s_{13}
  \}&=&0 \\
  \label{222}
  c_{12}s_{13} -  r  \frac{1+\lambda y}{1+\lambda x}
   s_{23}(c_{12} s_{23} s_{13}+s_{12} c_{23}
  )&=&0\\
  \label{333}
  s_{12}s_{13} -  r  \frac{1+\lambda y}{1+\lambda x}
   s_{23}( s_{12} s_{23} s_{13}-c_{12} c_{23}
  )&=&0.
  \end{eqnarray}

We find no non-trivial solution that satisfies this
set of equations. However, a limit of the second case to be 
considered next does yield a non CP violating mixing matrix and
reproduces the results given in Ref. \refcite{isdva}.

\subsection{{$\lambda y=1$} Case:}

  According to the definition,
  $\lambda y=1$ indicates that, if the ratio of  the
  numbers of  $\nu_\mu$ to $\bar\nu_\mu$
  is close to the ratio of the cross-sections of
  $\bar\nu_e$  to $\nu_e$, then this circumstance allows to
ignore the last term on the left hand side of Eq. (\ref{r0}).
 From Table 1 of Ref.
 \refcite{prd} we estimate $y \approx 2.06 \pm 0.31$,\footnote{It being the
value associated with the lowest atmospheric density in the
experiment, identified here as ``the top of the atmosphere.''}
while from Ref. \refcite{ggr} we infer $\lambda \approx 1/2.4$.
Thus, the required condition is fulfilled on ``accidental''
grounds. Further justification for ignoring the indicated term lies
in the fact that $J_{CP}$ is significantly suppressed by data-indicated
$U_{e3} \ll 1$. In any case $E$-dependent deviations from 
$\lambda y=1$ would contribute to departures from the exact $L/E$
flatness of the e-like event ratio. Similarly, we point out that
in certain range of $L/E$ the matter effects may become operative,
and these too would contribute to the indicated departure.

  Substituting the relevant elements of $U$ into
  Eqs. (\ref{f1}-\ref{f3}), similarly, we obtain

\begin{eqnarray}
  \label{aa1}
  c_{12} s_{12} c_{13}^2 +  \frac{2 r}{1+\lambda x} \{
  c_{12} s_{12} (s_{23}^2 s_{13}^2 - c_{23}^2)
  +(s_{12}^2-c_{12}^2) c_{23} s_{23} s_{13} c_{\delta}
  \}&=& 0 \\
  \label{aa2}
  c_{12}s_{13} -  \frac{2 r}{1+\lambda x}
   s_{23}(c_{12} s_{23} s_{13}+s_{12} c_{23}
  c_{\delta})&=&0\\
  \label{aa3}
  s_{12}s_{13} -  \frac{2 r}{1+\lambda x}
   s_{23}( s_{12} s_{23} s_{13}-c_{12} c_{23}
  c_{\delta})&=&0.
  \end{eqnarray}
{}From Eqs. (\ref{aa2},\ref{aa3}) we infer,
  \begin{equation}
  \label{bb1}
  s_{23}^2= \frac{1+\lambda x}{2 r}
  \end{equation}
  and
  \begin{equation}
  \label{bb2}
  c_{\delta}=0.
  \end{equation}
  So, the CP phase is $\pi/2$ or $-\pi/2$. Inserting 
Eqs. (\ref{bb1},\ref{bb2}) into Eq.  (\ref{aa1}),
  we have
  \begin{equation}
  \label{bb3}
  c_{23}^2=\frac{1+\lambda x}{2 r}
  \end{equation}
Finally, combining Eq. (\ref{bb1}) and Eq. (\ref{bb3}),
  we achieve the results:
  \begin{equation}
  \theta_{23}=\pi/4, \hspace{1cm}  r=1+\lambda x,
  \end{equation}
  That is, the mixing between the second and the
  third generations is maximal, and that
  the ratio of the numbers of $\nu_\mu$ to $\nu_e$
  equals to one plus the ratio of the numbers
  of $\bar\nu_e$  to $\nu_e$ events
  in case of no oscillations.

  As a result, the indicated L/E flatness in the
  the Super-Kamiokande data on the atmospheric neutrinos
   implies CP-violating maximal
  mixing matrix:

  \beq
  \label{Vpm}
  U^\pm=
  \left(
  \begin{array}{ccc}
  c_{12}\, c_{13} & s_{12}\, c_{13} & \mp i\, s_{13} \\
  -\frac{1}{\sqrt{2}}\left(s_{12} \pm i\, c_{12} \,s_{13}\right) &
  \frac{1}{\sqrt{2}}\left(c_{12} \mp i\, s_{12} \,s_{13}\right) &
  \frac{1}{\sqrt{2}} c_{13}\\
  \frac{1}{\sqrt{2}}\left(s_{12} \mp i\, c_{12}\, s_{13}\right) &
  -\frac{1}{\sqrt{2}}\left(c_{12} \pm i\, s_{12} \,s_{13}\right) &
  \frac{1}{\sqrt{2}} c_{13}
  \end{array}
  \right)
  \eeq
  where $U^+$ corresponds to $\delta_{13}=\pi/2$, and
  $U^-$ arises from $\delta_{13}=-\pi/2$. 

\section{Concluding Remarks}

\noindent
Corresponding to the two general
  forms for $U$, we obtain the following two measures of CP violation:
  \beq
  J_{CP}^\pm = \pm \frac{1}{2} c_{12} s_{12} c^2_{13} s_{13}
  = \pm \frac{1}{8} \sin\left(2\theta_{12}\right)
   \sin\left(2\theta_{13}\right) \cos\left(\theta_{13}\right)
  \label{jcp}
  \eeq
  In the limit $\theta_{13}$ vanishes
  the $U^\pm$ reduces to the result contained in Eq.  (26) of
  Ref. \refcite{isdva}, as it should. 
Preliminary indications that the $U$ matrix carries
  a similar form as given in Eq. (\ref{Vpm})
  can also be deciphered from a recent work
  of Barger, Geer, Raja, and Whisnant \refcite{bgrw}.
  Furthermore, for $\theta_{12}=\pi/4$, $U^+$ reads
\beq
 U^+=\left(
  \begin{array}{ccc}
{c_{13}}/{\sqrt{2}} & {c_{13}}/{\sqrt{2}} & - i s_{13}\\ 
-\left({1+i s_{13}}\right)/{2} & \left({1- i s_{13}}\right)/{2} 
&{c_{13}}/{\sqrt{2}} \\
\left({1-i s_{13}}\right)/{2} & -\left({1+i s_{13}}\right)/{2} 
& {c_{13}}/{\sqrt{2}}
\end{array}
\right)
\eeq	
which coincides with the Xing postulate \refcite{Xing}. The latter, 
originally  invoked to simultaneously allow for the a neutrino-oscillation
explanation of the atmospheric and solar neutrino data, turns out to be
dictated upon us by the indicated $L/E$ flatness.

  Since the CHOOZ experiment \refcite{chooz}
  constraints, for large-$\delta m^2$, $\sin^2\left(2\theta_{13}\right)$
  to be about $0.1$, even
  the large value of $\delta_{13} = \pm \pi/2$ implied by the present
  analysis, does not result in a maximal CP-violating difference:
  \beq
  \label{grav}
  P(\nu_{\alpha} \stackrel{L}{ \rightarrow} \nu_{\beta})
  -
  P(\bar\nu_{\alpha} \stackrel{L}{ \rightarrow} \bar\nu_{\beta})
  =
  4 J_{CP} \sum_{j<k}\sin\left(2\,\varphi_{jk}\right)
  \eeq
However, we  note that Eqs. (\ref{pp1},\ref{pp2})
  define a set of flavor-oscillation clocks, and these clocks must red-shift
  when introduced in a gravitational environment. If this
  environment is characterized by a
  dimensionless gravitational potential,  $\Phi_{grav}$, then
  in order that the flavor-oscillations suffer a gravitationally-induced
  red-shift
  we must replace, in
  Eq. (\ref{grav}),
  $\varphi_{jk}$ by $\left(1+\Phi_{grav}\right)\varphi_{jk}$.
  For other quantum-gravity effects on neutrino oscillations we refer
  the reader to Ref. \refcite{ggr}. Such gravitationally-induced modifications
  to a neutrino-sector CP violation may carry significant physical
  implications.

\section{Summary}

\noindent
  In summary, firstly, our discussion extended in this work
  seems to obligate us to accept a CP violated neutrino sector. And
  secondly, once CP is violated in neutrino system,
  the exact $L/E$ flatness of ${\cal R}_e$ implies that:
  (i) The mixing between the second and the third
  generations must be maximal,
  (ii) The ratio  ${\cal R}_e$ must be unity,
  (iii)  The CP-violating phase in the standard parameterization
  matrix is $\pi/2$ up to a sign ambiguity,
  (iv)   $N_{\nu_\mu} \sigma_{\nu_e}=N_{\bar
  \nu_\mu}\sigma_{\bar \nu_e}$, and
  finally that
  (v) $N_{\nu_\mu}/N_{\nu_e}=1+N_{\bar \nu_e}
  \sigma_{\bar \nu_e}/
  N_{ \nu_e} \sigma_{ \nu_e}$.

  Therefore, a dedicated study of the  ratio ${\cal R}_e$ in terms
  of its precise value, and its $L/E$ dependence, can become a powerful
  probe to study CP violation in the neutrino sector.
  Within the framework of this study, if the future
  data confirms  ${\cal R}_e$ to be unity
  for all zenith angles, then we must conclude that either
  there is no CP violation in the neutrino sector, or it is of the form
  predicted by equation (\ref{jcp}). This precise result, in conjunction
  with knowledge of $\theta_{12}$, $\theta_{13}$, and the associated
  mass-squared differences,  up to a sign ambiguity,
  completely determines the expectations for CP violation in all
  neutrino-oscillation channels.

However, the assumptions made in arriving the above results
may be violated to some extent, and we once again point out that 
the $E$-dependent deviations from 
$\lambda y=1$ would contribute to departures from the exact $L/E$
flatness of the e-like event ratio. Similarly, we note that
in certain range of $L/E$ the matter effects may become operative,
and these too would contribute to the indicated departure.
Once deviations from $\lambda y=1$ are fully
incorporated, the study of the $L/E$ flatness of the e-like event
ratio at Super-Kamiokande probes not only  
CP violation in the neutrino sector, but it also explores 
absence/presence of matter effects in atmospheric neutrino 
oscillations.  At present the available data contains significant
systematic and statistical errors, and, for that reason, these
higher order corrections are left to a future investigation.

\nonumsection{References}

\def\PR{{\bibit Phys. Rev.\/}$\;$}
\def\PRL{{\bibit Phys. Rev. Lett.\/}$\;$}
\def\PTP{{\bibit Prog. Theor. Phys.\/}$\;$}
\def\PL{{\bibit Phys. Lett.\/}$\;$}
\def\NP{{\bibit Nucl. Phys.\/}$\;$}
\def\MPL{{\bibit Mod. Phys. Lett. \/}$\;$}

\end{document}